# Universal Deoxidation of Semiconductor Substrates Assisted by Machine-Learning and Real-Time-Feedback-Control

Chao Shen,[§,†] Wenkang Zhan,[‡,†] Jian Tang,[∥] Zhaofeng Wu,*[,§] Bo Xu,[‡,†] Chao Zhao,*[,‡,†] and Zhanguo Wang[‡,†]

[§] School of Physics Science and Technology, Xinjiang University, Urumqi, Xinjiang 830046, China

[‡] College of Materials Science and Opto-Electronic Technology, University of Chinese Academy of Science, Beijing 101804, China

[†] Key Laboratory of Semiconductor Materials Science, Institute of Semiconductors, Chinese Academy of Sciences & Beijing Key Laboratory of Low Dimensional Semiconductor Materials and Devices, Beijing 100083, China

[∥] School of Physical and Electronic Engineering, Yancheng Teachers University, Yancheng 224002, China

ABSTRACT

Thin film deposition is an essential step in the semiconductor process. During preparation or loading, the substrate is exposed to the air unavoidably, which has motivated studies of the process control to remove the surface oxide before thin film deposition. Optimizing the deoxidation process in molecular beam epitaxy (MBE) for a random substrate is a multidimensional challenge and sometimes controversial. Due to variations in semiconductor materials and growth processes, the determination of substrate deoxidation temperature is highly dependent on the grower's expertise; the same substrate may yield inconsistent results when evaluated by different growers. Here, we employ a machine learning (ML) hybrid convolution and vision transformer (CNN-ViT) model. This model utilizes reflection high-energy electron diffraction (RHEED) video as input to determine the deoxidation status of the substrate as output, enabling automated substrate deoxidation under a controlled architecture. This also extends to the successful application of deoxidation processes on other substrates. Furthermore, we showcase the potential of models trained on data from a single MBE equipment to achieve high-accuracy deployment on other equipment. In contrast to traditional methods, our approach holds exceptional practical value. It standardizes deoxidation temperatures across various equipment and substrate materials, advancing the standardization research process in semiconductor preparation, a significant milestone in thin film growth technology. The concepts and methods demonstrated in this work are anticipated to revolutionize semiconductor manufacturing in optoelectronics and microelectronics industries by applying them to diverse material growth processes.

**Introduction**

Epitaxy thin film is the heart of state-of-the-art optoelectronic and microelectronic devices. These layers' crystal quality and defect density are greatly affected by growth conditions and the starting surface after substrate preparation. The deoxidation process of semiconductor substrates is critical for the growth of high-quality epitaxial layers by molecular beam epitaxy, metal-organic vapor phase epitaxy (MOVPE), etc.[1] Although etchants are usually used to intentionally remove the oxide layer before epitaxy, a fresh natural oxide layer instantly forms after exposure to the ambient atmosphere.[2] Deoxidizing the substrate in the vacuum chamber before the epitaxy is necessary.[3] Thermal annealing is a general way of obtaining an epi-ready surface on the substrates for the following growth.[4] However, the deoxidation duration and temperature depend on the oxide thickness and structure of substrates, which can be complicated and controversial.[1, 5, 6] It is needed to prevent further surface damage during thermal deoxidation of substrates. Otherwise, it will compromise the quality of the following grown epilayers.[7, 8] It is reported that the reproductivity of AlGaInAs-based lasers can be improved with proper substrate deoxidation.[9]

A reflection high-energy electron diffraction (RHEED) is typically used to monitor the surface reconstruction to determine the desorption of oxides from substrates, which is implemented by heating slowly and carefully by experienced growers for long periods.[10-12] The diffraction patterns with dynamic and overlapping information are also challenging to interpret. Artificial intelligence (AI) was used to analyze RHEED patterns during MBE growth. It detects RHEED patterns in real-time during the Si(111) substrates deoxidation process and classifies them by their similarity to specific surface reconstruction.[13] Peter R. Wiecha used deep-learning-based RHEED image-sequence classification to identify the exact deoxidation moment.[14] We developed a machine learning (ML) model trained using RHEED videos as input and provided real-time feedback on surface morphologies for process control.[15] However, previous reports have typically concentrated on specific materials and the post-analysis of model results for collected data, neglecting the construction of universally applicable models for similar applications that enable real-time deployment. The dataset collection and application were limited to the same MBE equipment, with no endeavor to explore models capable of maintaining consistent performance across different equipment. If a universal model can be achieved, it can reduce reliance on grower experience, improve the reproducibility of material preparation, and better adapt to constantly evolving technology and material advancements.

In this paper, we amassed a substantial volume of deoxidation RHEED video data encompassing GaAs, along with a smaller dataset involving Ge and InAs. We successfully developed a highly robust hybrid Convolutional and Visual Transformer (CNN-ViT) model through numerous optimizations of model training parameters. This model demonstrates adaptability to samples of varying resolutions as input, with its performance unaffected by camera hardware resolution constraints. A detailed analysis of the model parameters revealed a distinct boundary in the classification results output by the model, signifying its high sensitivity to data. Furthermore, the region with the highest output weight from the CNN-ViT model's attention module aligns with the region of interest for experienced growers. This consistency indicates that the model is reasonable and possesses strong interpretability. In-situ automatic deoxidation experiments were conducted, and during the dynamic substrate heating process, the model accurately identified the deoxidation status of GaAs, Ge, and InAs substrates, providing precise deoxidation temperatures.

Additionally, the model was applied to analyze RHEED data collected from substrates and equipment not included in the datasets, exhibiting exceptional accuracy and underscoring its robust universal performance. This study demonstrated that a single data source can create universal models across various devices in different material systems. Additionally, the universality of the model enables the standardization of each stage of material growth, mitigating errors caused by traditional human experience-based judgments. The future development of more universal standardized models is expected further to advance the standardization process in the semiconductor manufacturing.

**Experimental**

The samples were prepared using the Riber 32P MBE system, which was equipped with an arsenic (As) valve cracker and effusion cells. The substrate temperature was measured using a C-type thermocouple. Before introducing the substrate into the growth chamber for deoxidation, a 6-hour low-temperature degassing was conducted in the buffer chamber. RHEED in the MBE growth chamber facilitated the analysis and monitoring of the substrate surface during the deoxidation process. RHEED patterns were recorded at 12 kV electron energy (RHEED 12, from STAIB). A darkroom, equipped with a camera, was positioned to continuously capture RHEED videos while rotating the substrate at 20 rpm. The exposure time was 100 ms, and the frame sampling rate was 8 frames per second (fps). As shown in Figure 1, the data captured by the camera is processed to preserve only the selected square matrix area during the dynamic heating and deoxidation process. Subsequently, the collected data is segmented into various images in chronological order. Each image undergoes normalization on the brightness channel and is converted into a two-dimensional matrix with an 8-bit depth. These multiple continuous two-dimensional matrices are then connected and combined to form a new three-dimensional matrix, serving as the input for the model. The output of the model determines whether the substrate is deoxidized. If deoxidation is incomplete, the substrate temperature continues to rise. If deoxidation is complete, the current deoxidation temperature is obtained, signifying the experiment's conclusion. By preprocessing the raw RHEED video data, we obtained 320,000 NPY files for training (see Supplementary Information for the classification criteria for data, S1).

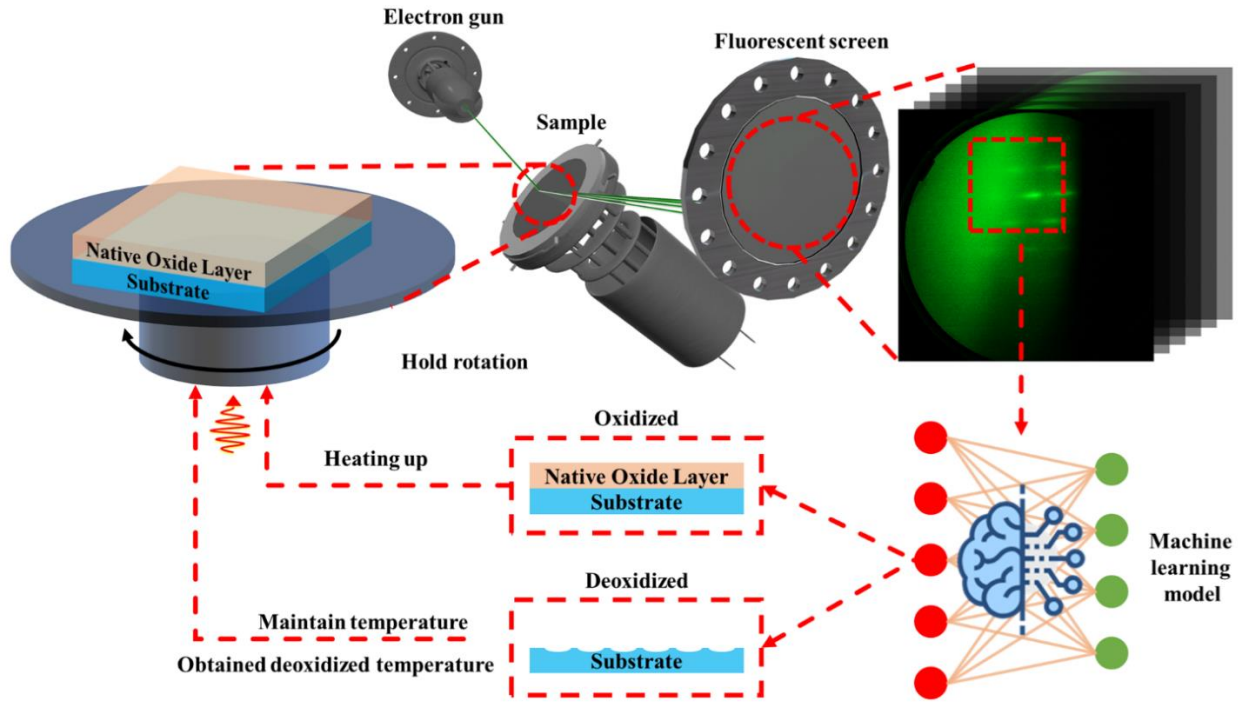

Figure 1. The overall framework of the experiment.

CNN-ViT model has become our preferred choice, given sufficient training data.[16] Compared to other convolutional-based models, CNN-ViT incorporates the Transformer's self-attention mechanism, enhancing the model's ability to capture global information from images.[17, 18] This reduces the dependence on a large number of parameters while ensuring robustness.[19, 20] Traditional convolutional architectures might be limited by receptive fields when processing global information, often requiring more parameters to capture a broad range of global information.[21-23] Moreover, ViT is not sensitive to input position and is more flexible with images of different sizes.[24] Our model also includes an upsampling layer before the convolutional layer, enhancing adaptability to inputs of varying pixel sizes without altering the model structure, as shown in Figure 3a.[25, 26] Additionally, the Transformer structure supports parallel computing during the training and deployment processes, providing an advantage in the model deployment process.[20] This allows for more comprehensive utilization of the collected raw data. The raw data processed by the method in Figure 3a will be standardized into a fixed-size block matrix.[27] The position of each block will be encoded and embedded into the image matrix, forming an input sequence. This sequence is then input into the Transformer encoder, where each layer encompasses a multi-head attention mechanism and a feedforward neural network, as illustrated in Figure 3b.[28, 29] Finally, the outcomes processed by the feedforward neural network are consolidated and sequentially output through the Multilayer Perceptron (MLP) layer, GELU layer, and another MLP layer. In Figure 3c, the input sequence of the multi-head attention mechanism is linearly transformed and divided into multiple subspaces, each termed an attention head—such as value (V), key (K), and query (Q).[30] Each attention head possesses its weight matrix for computing the attention distribution.[17] Throughout the data processing, the output of each attention head is consolidated and subsequently linearly transformed to yield the final output of the multi-head attention mechanism.

We conducted tests to assess the model's accuracy under varying parameters and input sizes (see Supplementary Information for the optimization of model parameters, S2). As the depth and head numbers progressively increase, the validation accuracy demonstrates an upward trend, while the validation loss shows a downward trend, as shown in Figure 2d. However, setting the depth and head to exceed 16 does not significantly improve validation accuracy, and the validation loss does not exhibit a substantial decrease. This suggests that further increases in these parameters only result in elevated model complexity without effectively enhancing model performance.

Additionally, as the number of images increases, the model's performance also exhibits an upward trend, as shown in Figure 2e. However, the model's accuracy decreases when the number of images exceeds 12. This phenomenon may arise due to the high complexity of the parameters, necessitating more epochs for the model to yield improved results. Nevertheless, to balance training time and accuracy, selecting every 12 images as input for the model proves to be the optimal choice. We chose this approach because rotating the substrate at 20 revolutions per minute and a frame sampling rate of 8 frames per second results in 24 frames of RHEED data collected during a single substrate rotation. Among these 24 frames, 12 contain duplicate information. Inputting 12 frames of images each time effectively avoids duplicate images in the RHEED-collected data, preventing data redundancy and enhancing the efficiency of model data processing.

Finally, we also adjusted the pixel count of each image to explore changes in model accuracy. After training these models for 100 epochs, we observed that as the image pixels increased, the performance of the models gradually improved, as depicted in Figure 2f. However, when the resolution exceeds 64, the improvement in model accuracy becomes limited. With an input pixel size of 128, the deployed model in the program can generate approximately 9 results per second-a value very close to the camera sampling rate, fully utilizing the RHEED data collected by the camera. Furthermore, considering that richer input information can enhance the model's accuracy, we have ultimately chosen to set the input pixel size for each image to 128. In summary, we have determined the model structure, and after sufficient training, the validation accuracy of the model can reach 99.95%, with an average validation loss of only 0.001646399.

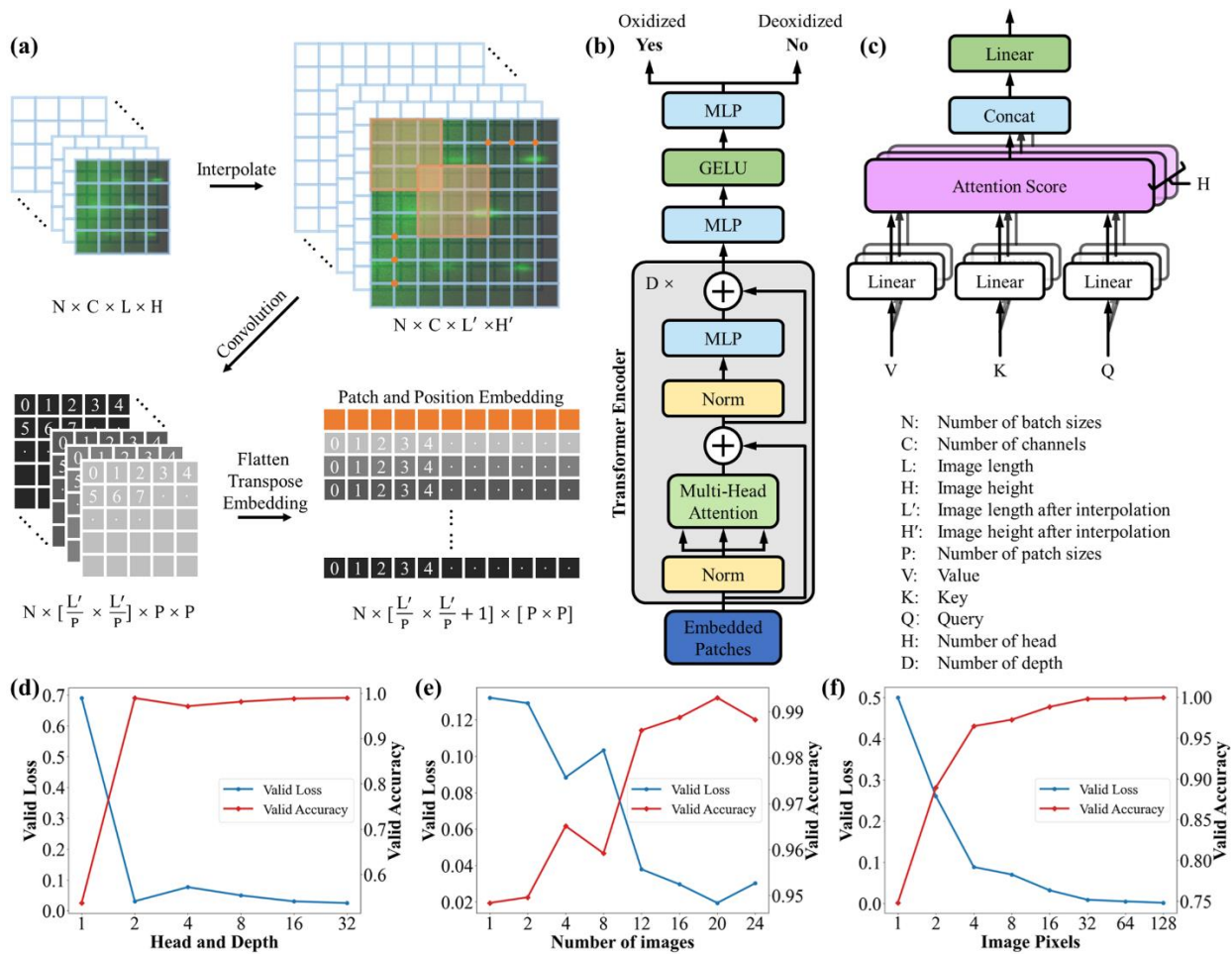

Figure 2. CNN-ViT model structure: a) Schematic diagram illustrating convolutional data processing with upsampling. b) The architecture of the ViT. c) Multi-head attention mechanism. The variation of model validation accuracy and validation loss under different (d) heads and depths, (e) image numbers, and (f) image pixels.

We selected typical deoxidation data to analyze the features learned by the CNN-ViT model, as illustrated in Figure 5a. Firstly, we perturbed the original data slightly, generated corresponding adversarial samples, and observed the model's response. It was found that the generated samples were nearly identical to the original image, indicating the model's good robustness, as shown in Figure 5b.[31, 32] Subsequently, we visualized the regions the model focused on in the input data and generated an attention heatmap, as shown in Figure 5c.[33, 34] Figure 5a was annotated based on the grid partitioning method, highlighting regions of interest in the attention heatmap. The attention area in the heatmap was observed to be focused near the specular spot of RHEED, which aligns with the experienced grower judgment process based on the brightness difference between the specular spot and its surrounding background, demonstrating strong interpretability.[35, 36] Next, we randomly selected 5 oxidized data and 5 deoxidized data, using the t-distributed Stochastic Neighbor Embedding (t-SNE) algorithm to map the high-dimensional features of the model to a two-dimensional space, forming a scatter plot, as shown in Figure 5d.[37] The scatter plot exhibited clear separation into two categories with a distinct boundary, as indicated by the red dashed box in Figure 5d, implying that the model possesses good sensitivity in determining the substrate deoxidation state.[38, 39] Additionally, the activation values of each feature map on the training set were averaged and plotted as curves, as shown in Figure 5e. We selected three typical convolutional kernels and the feature maps output by the convolutional layers from Figure 5e and visualized their parameters in Figures 5f and 5g. The weight changes of the convolutional kernel and convolutional layer output near the RHEED specular spot were significant, once again indicating the consistency between the models and human discrimination methods. Subsequently, we attempted to analyze data collected from other MBE equipment for the model, as shown in Figure 5h. It was observed that even when the model was applied to substrates at different rotation speeds or to GaSb substrate data from non-datasets, the average probability of identifying deoxidized was as high as 96.3%. The probability of identifying oxidized was as high as 88.9%, indicating that the model exhibits strong versatility even without undergoing a new round of training.

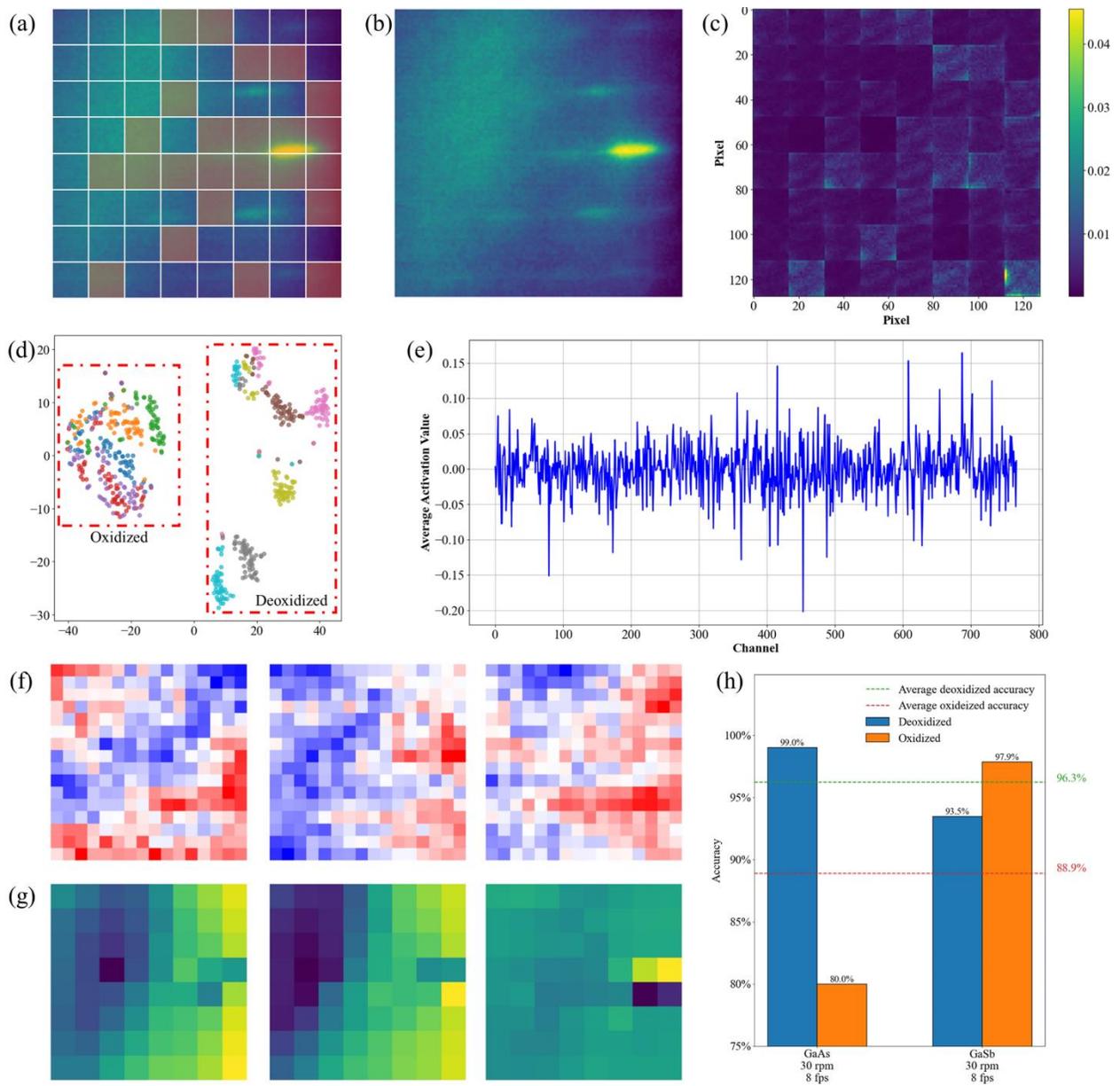

Figure 3. Model Feature Analysis: a) Visualization of original images. b) Generation of adversarial samples. c) Attention heatmap. d) t-SNE visualization of high-dimensional features. e) Average Activation Curve. f) Visualization of convolutional kernels. g) Visualization of feature maps after convolution. h) Model processing results of data on non-datasets.

The complete removal of the substrate oxide layer is crucial to ensure that the quality of the bulk semiconductor substrate is not compromised.[5] Additionally, deoxidation is generally intended to be conducted at the lowest possible temperature to minimize non-stoichiometric effects caused by inconsistent evaporation of atoms on the substrate surface.[5] Therefore, when substrate oxidation is detected, the program will increase the temperature above the current level for the substrate (see Supplementary Information for the model and program development environment, S3). Subsequently, the model maintained this temperature for a period before making a secondary judgment, repeating the process until it judged that the substrate had been deoxidized. Furthermore, the criterion for substrate deoxidation is established by having 95% of 24 consecutive judgment of the model yield a deoxidation outcome. This avoids situations where inaccurate deoxidation temperature recognition may occur due to the uneven thickness of the substrate surface oxide.

**Results and discussion**

We conducted automatic deoxidation experiments on GaAs substrates using the program. The program operation stages were divided based on changes in "Reminder Information" on the program interface, as illustrated in Figure 4a. 11 heating cycles were performed throughout the program operation, starting from 350 °C and reaching 405 °C for deoxidation, as depicted in Figure 4b. After each heating, the substrate will be kept at this temperature for 6 minutes. Before the end of 6 minutes, the RHEED shutter was opened, and the model was employed for continuous judgement. The judgement results of the model were plotted using a scatter plot, and the moving average method was used for statistical analysis of the judgment results, as shown in Figure 4c. At the deoxidation temperature, the probability of the model output "Yes" increased rapidly, while at other temperature points, the model output results mainly showed "No". The program continuously counted 24 output results of the model, and only when the probability of the model output being "Yes" exceeded 95%, indicating that the RHEED obtained from any angle of the substrate could be judged as deoxidation, was the current temperature considered the deoxidation temperature, as shown in Figure 4d. We selected typical RHEED images from 350 ℃, 400 ℃, and 405 ℃, as shown in Figures 4e-4g. The RHEED images at 350 ℃ and 400 ℃ revealed relatively blurry bright spot features with weak brightness. However, the RHEED pattern obtained at 405 ℃ exhibited a clear and distinct light spot outline, indicating successful substrate deoxidation at 405 ℃. The program achieved automatic deoxidation of GaAs substrates. In addition, we successfully conducted similar experiments on Ge and InAs substrates (see Supplementary Information for the automatic deoxidation experiment on Ge substrate and the automatic deoxidation experiment on InAs substrate, S4 and S5).

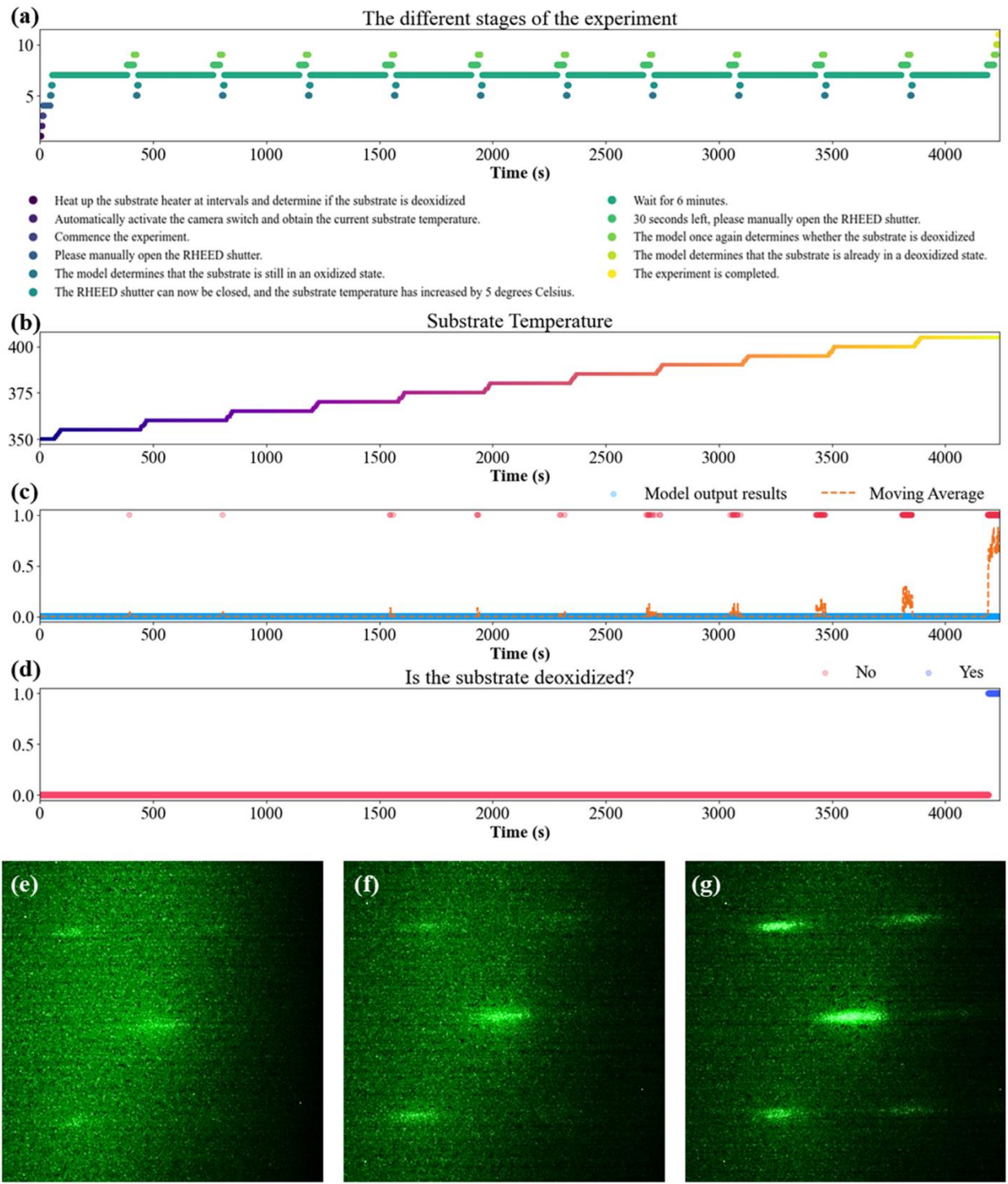

Figure 4. Automatic deoxidation experiment on GaAs substrate. a) Division of program running stages. b) Substrate temperature curve. c) The output results of the model and the statistical results of the moving average method. d) The program determines whether the substrate is in the deoxidation stage. e) RHEED captured at the 350 ℃. f) RHEED captured at the 400 ℃. g) RHEED captured at 405 ℃.

Moreover, we recorded a section of RHEED data of deoxidation on a GaSb substrate from the Riber C21 MBE system, with a substrate rotation speed of 30 rps and a camera sampling rate of 8 fps and submitted the data to the model for processing. The output results of the model are shown in Figure 5a. It can be observed that from the $400^{th}$ sequence to the $800^{th}$ sequence, the probability of the model outputting "Yes" gradually increases, and the statistical curve of the moving average method steadily increases. A typical RHEED was achieved from the $400^{th}$, $600^{th}$, and $800^{th}$ sequences, as shown in Figures 5b-5d. At the $400^{th}$ sequence, there were almost no patterns in the RHEED pattern, indicating that deoxidation had not occurred. However, at the $600^{th}$ sequence, sharp main light spots could be seen in the RHEED, but if the surrounding light spots were not obvious, it indicated that the substrate had gradually approached the deoxidation state. In the $800^{th}$ sequence, the specular spot in the RHEED pattern not only becomes sharper but also more prominent in brightness, and the surrounding small spots gradually become prominent, indicating that the substrate has been deoxidized. The data results confirm that the model can identify the deoxidation of unknown materials for other devices and has strong universality.

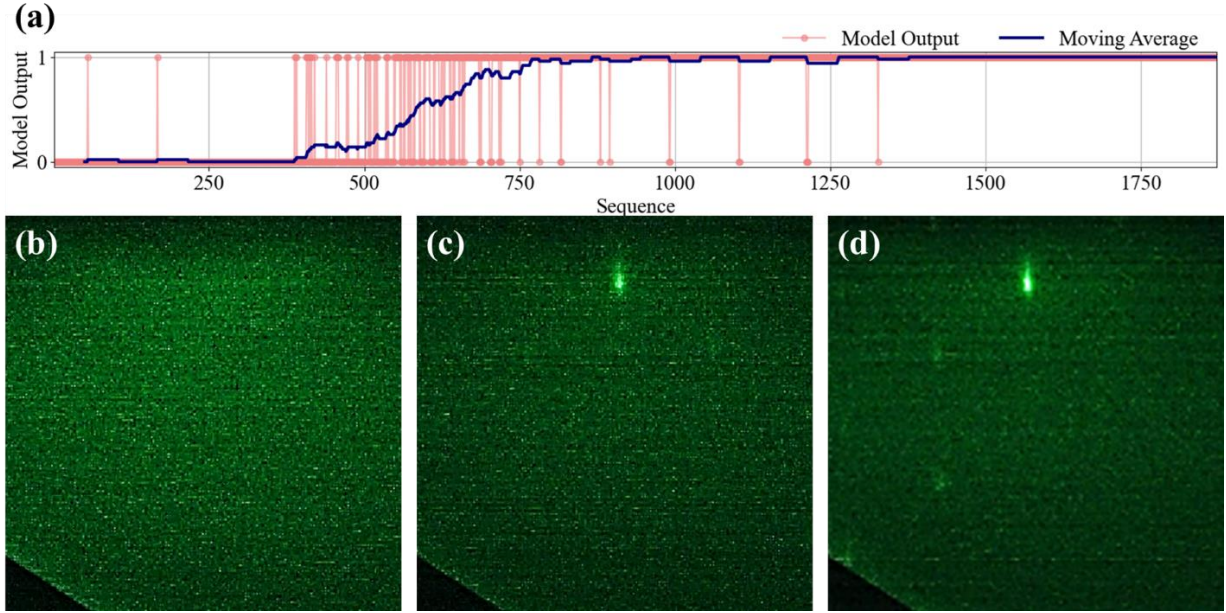

Figure 5. Model results of GaSb substrate deoxidation data. a) The output results of the model and the statistical results of the moving average method. b) RHEED captured at sequence 400. c) RHEED captured at sequence 600. d) RHEED captured at sequence 800.

**Conclusions**

In this report, we comprehensively explore the automatic deoxidation of substrates using a hybrid CNN-ViT model. The model is trained on diverse datasets containing deoxidation RHEED video data from GaAs, Ge, and InAs substrates, exhibiting significant adaptability to various resolutions and camera hardware. The detailed analysis of model parameters, attention mechanisms, and features emphasizes its robustness and consistency with human empirical methods, demonstrating strong interpretability. In addition, we conducted in-situ automatic deoxidation experiments. The model accurately identifies the deoxidation state of GaAs, Ge, and InAs substrates during the dynamic substrate heating process, providing accurate deoxidation temperature. The model demonstrates considerable accuracy when processing GaSb substrate data from different MBE equipment. The universality of this model on various equipment and substrates provides a way to promote the standardization process in the semiconductor manufacturing field, and the development of more universal standardized models is a trend for the future.

## ASSOCIATED CONTENT

**Supporting Information**. A listing of the contents of each file supplied as Supporting Information should be included. For instructions on what should be included in the Supporting Information as well as how to prepare this material for publications, refer to the journal's Instructions for Authors.

## AUTHOR INFORMATION

**Corresponding Author**

*Email: zhaochao@semi.ac.cn, wuzf@xju.edu.cn

**Author Contributions**

C. S., W. K. Z., and J. T. contributed equally. C. Z. conceived of the idea, designed the investigations and the growth experiments. C. S. and W. K. Z. performed the molecular beam epitaxial growth. C. S. did the sample characterization. C. S., C. Z., J. T., and Z. F. W wrote the manuscript. C. Z. led the molecular beam epitaxy program. B. X. and Z. G. W. supervised the team. All authors have read, contributed to, and approved the final version of the manuscript.

## ACKNOWLEDGMENT

This work was supported by the National Key R&D Program of China (Grant No. 2021YFB2206503), National Natural Science Foundation of China (Grant No. 62274159), the "Strategic Priority Research Program" of the Chinese Academy of Sciences (Grant No. XDB43010102), and CAS Project for Young Scientists in Basic Research (Grant No. YSBR-056).